\documentclass[12pt,a4]{article}
\usepackage{cite}
\usepackage{amsmath}
\usepackage{subcaption}
\usepackage{amssymb}
\usepackage{amsfonts}
\usepackage{array}
\usepackage{graphicx}
\usepackage{mathrsfs}
\usepackage{multirow}
\usepackage{siunitx}
\newcommand{\bea}{\begin{eqnarray*}}
\newcommand{\eea}{\end{eqnarray*}}
\newcommand{\beao}{\begin{eqnarray}}
\newcommand{\eeao}{\end{eqnarray}}

\begin{document}
\title{Minimal Length Phenomenology and the Black Body Radiation}

\author{Pasquale Bosso\footnote{pasquale.bosso@uleth.ca}, Juan Manuel L{\'o}pez Vega\footnote{jm.lopez@uleth.ca}\\ \\
University of Lethbridge,\\
4401 University Drive, Lethbridge, Alberta, Canada, T1K 3M4}
\date{}
\maketitle

\begin{abstract}
The generalized uncertainty principle (GUP) modifies the uncertainty relation between momentum and position giving room for a minimal length, as predicted by candidates theories of quantum gravity. Inspired by GUP, Planck's distribution is derived by considering a new quantization of the electromagnetic field. We elaborate on the thermodynamics of the black body radiation obtaining Wien's law and the Stefan-Boltzmann law. We show that such thermodynamics laws are modified at Planck-scale.
\end{abstract}

\section{Introduction}
Candidate theories of quantum gravity, such as string theory, loop quantum gravity, as well as \textit{gedanken} experiments in black hole physics predict the existence of a minimal uncertainty in the position \cite{gross1987high,amati1987superstring,gross1988string,amati1989can,konishi1990minimum,garay1995quantum,adler1999gravity,scardigli1999generalized,Capozziello:1999wx}. Such a minimal length is in direct contradiction with the Heisenberg uncertainty principle. Thus, a modification to the Heisenberg principle has to be introduced that is expected to be relevant at the Planck scale. Such a modification is considered in phenomenological models of quantum gravity \cite{Amelino-Camelia:2008aez}. In particular, the Generalized Uncertainty Principle (GUP) consists in modifying the standard position-momentum commutation relation by including a function of the momentum operator \cite{maggiore1993generalized,maggiore1993algebraic,kempf1995hilbert,das2008universality,jizba2010uncertainty,ali2011proposal,frassino2012casimir,husain2013generalized,scardigli2017gup,bosso2017planck,scardigli2018modified,luciano2019gup,blasone2020heuristic,bosso2021quasi}. A typical model involves a quadratic modification of the form

\begin{equation}
[q,p]=i\hbar[1+\gamma^{2}p^{2}]\label{eq1},    
\end{equation}

where 

\begin{equation}
    \gamma=\frac{\gamma_{_{0}}}{M_{_{pl}}c},
\end{equation}

and $\gamma_{_{0}}$ is a dimensionless parameter that determines the energy scale at which such a modification takes place. Such a parameter is determined experimentally. For example, in the case of a macroscopic harmonic oscillator, it has been possible to find an upper bound for such a parameter of the order of $10^{4}$  \cite{Bushev:2019zvw}. If it is assumed to be of order unity, the modification will be relevant at the Planck energy.

This model has been applied to several low-energy systems searching for indirect quantum gravity effects. Examples of such indirect tests concern the quantum harmonic oscillator, condensed matter, and atomic experiments \cite{hamma2011background,dos2013toward,bosso2017planck,danshita2017creating}. Further studies have been pursued in statistical mechanics \cite{shababi2020non,el2020some,haine2021searching,das2021test}, where the GUP affects thermodynamic variables.

The black body radiation represents one of the most interesting problems in the history of physics. The solution to the black body problem with the introduction by Planck of the hypothesis of the quantized energy exchanges revealed the issues of classical mechanics and laid the foundation for the development of modern quantum mechanics. Specifically, Planck assumed that the energy exchange between the modes of radiation enclosed in a cavity is proportional to the frequency of the mode, $\Delta  E=h\nu$. That means that the energy is not absorbed continuously, but discretely. Such a discretization is related to the discrete energy spectrum of a quantum harmonic oscillator; the energy difference between two neighbouring energy levels is proportional to the frequency of the oscillator. However, the introduction of a minimal length changes such energy differences \cite{kempf1995hilbert,bosso2017planck,bosso2021generalized}. Thus, Eq.\eqref{eq1} implies a change in Planck's postulate. In this paper, we elaborate on the modification of black body thermodynamics. By introducing a new modification on the radiation field inspired by the GUP, we intend to study Planck's distribution and the corresponding GUP modification. Studying the resulting expression, we obtain further features of the black body distribution, such as Wien's law and the Stefan-Boltzmann law. Similar considerations have been elaborated in \cite{maziashvili2011corrections,Nozari2012BlackBR,Shababi:2013kua} following different techniques. However, the novelty of the present approach consists in following Bose's statistical method \cite{bose1994planck} when studying the statistical properties of a photon gas. 

The paper is organized as follows. In section 2, we introduce the quantization relation for the radiation field inspired by GUP. In section 3, we elaborate a statistical analysis using the modified spectrum to obtain Planck's distribution. In sections 4 and 5, we obtain the modified Wien's law and Stefan-Boltzmann law to complete the study of the black body radiation. In section 6, we conclude by presenting future perspectives.

\section{GUP modification to the radiation field}

In this section, we review the GUP modification to the electromagnetic field quantization. According to \cite{bosso2018potential}, such a modification is introduced by modifying the generalized coordinates and momenta of the electromagnetic field $\textbf{q}_{\textbf{k}}$ and $\textbf{p}_{\textbf{k}}$, respectively, as follows

\begin{align}
[\textbf{q}_{\textbf{k}},\textbf{p}_{\textbf{k'}}]=i\hbar\delta_{\textbf{k,k'}}[1+\gamma_{_{EM}}^{2}\textbf{p}_{\textbf{k}}^{2}],  
\end{align}

where

\begin{align}
\gamma_{_{EM}}=\frac{\gamma_{_{0}}}{\sqrt{M_{_{pl}}}c}.
\end{align}

With this modification, the energy spectrum with GUP for a mode with wave vector \textbf{k} is \cite{bosso2017planck}

\begin{align}
E^{k}_{n}=\hbar\omega^{k}\bigg\{\bigg(n+\frac{1}{2}\bigg)+\frac{\hbar\omega^{k}  }{4}\gamma_{_{EM}}^{2}(1+2n+2n^{2})\bigg\},   
\end{align}

where $k$ is the magnitude of the wavenumber used to label the different modes. Here, we see that, differently from the standard theory, the energy difference between two neighbouring levels, $n$ and $n+1$, depends on the level $n$

\begin{equation}
\Delta E_{n}^{k}=\hbar\omega^{k}+\beta^{k}(n+1)\label{eq7}, 
\end{equation}

where $\beta^{k}=(\hbar\omega^{k}\gamma_{_{EM}})^{2}$. This term carries the modification to the energy due to GUP. It is worth observing that in the limit $\gamma_{_{EM}}\rightarrow 0$, we recover the usual expression for the energy of a photon.

\section{Modified Planck's Law}

Here, we follow the argument introduced by Bose to derive Planck's distribution  
 \cite{bose1994planck}. Considering the statistics of indistinguishable photons, we construct a distribution $Z_{j}^{k}$ for the number of phase space cells with $j$ of quanta in a particular frequency range $d\omega^{k}$.

By introducing the new quantum of energy Eq.$\eqref{eq7}$, the total energy is then 

\begin{equation}\label{eq8}
\begin{split}
E&=\sum_{k}(\Delta E_{1}^{k}Z_{1}^{k}+2\Delta E_{2}^{k}Z_{2}^{k}+3\Delta E_{3}^{k}Z_{3}^{k}+...)\\ &=\sum_{k}\left[\hbar\omega^{k}N^{k}+\beta^{k}(N^{k}+M^{k})\right],
\end{split}
\end{equation}

where
\begin{align}
& N^{k}=\sum_{j}jZ_{j}^{k},
& M^{k}=\sum_{j}j^{2}Z_{j}^{k}\label{eq9}.
\end{align}

For any distribution $Z_{j}^{k}$, the total number of cells available for the system in a range $d\omega^{k}$ is $A^{k}$, expressed by 

\begin{align}
A^{k}=\sum_{j}Z_{j}^{k}\label{eq10}.   
\end{align}

The probability of observing a particular distribution $Z_{j}^{k}$ is related to the number of different ways that particular distribution can be formed, that is, the permutation of occupied cells 

\begin{align}
    W=\prod_{k}\frac{A^{k}!}{\prod_{j}Z_{j}^{k}!}.
\end{align}

For a large number of photons, we adopt the following expression using Stirling's approximation

\begin{align}
\log W=\sum_{k}A^{k}\log A^{k}-\sum_{k}\sum_{j}Z_{j}^{k}\log Z_{j}^{k}\label{eq11}.  
\end{align}

We note that the system is subject to two constraints i.e. Eq.\eqref{eq8} and Eq.\eqref{eq10}. We determine the distribution $\bar{Z}_{j}^{k}$ that maximizes $W$ and keeps $E$ and $A^{k}$ constant

\begin{align}
&\delta \left\{\log W+\lambda\sum_{k}\sum_{j}Z_{j}^{k}+\gamma\sum_{k}\left[\hbar\omega^{k}N^{k}+\beta^{k}(N^{k}+M^{k})\right]\right\}=0,
\end{align}

where $\lambda$ and $\gamma$ are Lagrange multipliers. The variation with respect to $Z_{j}^{k}$ gives

\begin{align}
\sum_{k}\sum_{j}\delta Z_{j}^{k}\left\{1+\log Z_{j}^{k}+\lambda+\gamma\left[j\hbar\omega^{k}+\beta^{k}j(1+j)\right]\right\}=0.    
\end{align}

As each $\delta Z_{j}^{k}$ is arbitrary, we can impose the curly brackets to vanish obtaining

\begin{align}
\bar{Z}_{j}^{k}=B^{k}e^{-\gamma\hbar\omega^{k}j}e^{-\gamma\beta^{k}j(j+1)}\label{eq13},
\end{align}

where
\begin{equation}
B^{k}=e^{-1-\lambda^{k}}.    
\end{equation}

The second exponential in Eq.\eqref{eq13} carries the contribution due to GUP. For values of the temperature much smaller than the Planck temperature, \(T_{P}=1.41\times 10^{32}\ \text{K}\), and therefore out of the Planck scale, we can approximate Eq.\eqref{eq13} writing 

\begin{equation}
\bar{Z}_{j}^{k}=B^{k}e^{-\gamma\hbar\omega^{k}j}\left[1-\gamma\beta^{k}j(j+1)\right]\label{eq15}.    
\end{equation}

Due to such approximation, there is a maximum value for \(j^{k}\). Exceeding such value 

\begin{equation}
    j_{_{max}}^{k}=\left\lfloor\bigg(\frac{1}{\gamma\beta^{k}}+\frac{1}{4}\bigg)^{1/2}-\frac{1}{2}\right\rfloor,
\end{equation}

the approximation is no longer valid. In turn, such a maximum value for $j^{k}$ corresponds to a maximum value for the energy

\begin{equation}
E^{k}_{_{max}}=\hbar\omega^{k}+\beta^{k}\bigg\{\bigg(\frac{1}{\gamma\beta^{k}}+\frac{1}{4}\bigg)^{1/2}+\frac{1}{2}\bigg\}.
\end{equation}

Having the most probable distribution  Eq.\eqref{eq15}, we can proceed following Bose's approach \cite{bose1994planck}.
We calculate Eq.\eqref{eq9} and Eq.\eqref{eq10} using the geometric series and its derivatives. For the total number of cells, we obtain the following expression 

\begin{equation}
\begin{split}
  &A^{k}=\sum_{j}\bar{Z}_{j}^{k}\\
 &=\frac{B^{k}}{(1-e^{-\gamma\hbar\omega^{k}})^{3}}\left[(1-e^{-\gamma\hbar\omega^{k}})^{2}-2\gamma\beta^{k}e^{-\gamma\hbar\omega^{k}}\right]\label{a27},
 \end{split}
\end{equation}

which can be solved for $B^{k}$. Then, substituting in Eq.\eqref{eq9}, we find

\begin{equation}
\begin{split}     
&N^{k}=\sum_{j}j\bar{Z}_{j}^{k}\\
& =\frac{A^{k}}{e^{\gamma\hbar\omega^{k}}-1}\left\{\frac{(1-e^{-\gamma\hbar\omega^{k}})^{2}-2\gamma\beta^{k}(1+2e^{-\gamma\hbar\omega^{k}})}{(1-e^{-\gamma\hbar\omega^{k}})^{2}-2\gamma\beta^{k}e^{-\gamma\hbar\omega^{k}}}\right\},
\end{split}
\end{equation}

\begin{equation}
\begin{split}
 &M^{k}=\sum_{j}j^{2}\bar{Z}_{j}^{k}\\
& =\frac{A^{k}e^{-\gamma\hbar\omega^{k}}}{(1-e^{-\gamma\hbar\omega^{k}})^{2}}\left\{\frac{(1-e^{-\gamma\hbar\omega^{k}})^{2}(1+e^{-\gamma\hbar\omega^{k}})-2\gamma\beta^{k}(1+7e^{-\gamma\hbar\omega^{k}}+4e^{-2\gamma\hbar\omega^{k}})}{(1-e^{-\gamma\hbar\omega^{k}})^{2}-2\gamma\beta^{k}e^{-\gamma\hbar\omega^{k}}}\right\}.
\end{split}
\end{equation}

The number of modes per unit volume in the black body cavity for a particular frequency range $d\omega^{k}$ is given by the Rayleigh-Jeans expression 

\begin{equation}
\Re (\omega) d\omega^{k}=\frac{(\omega^{k})^{2}}{\pi^{2}c^{3}}d\omega^{k}\label{rj}.   
\end{equation}

This expression is obtained using geometrical arguments \cite{beale2011statistical}. As such expression is purely geometrical, it is not affected by the GUP modification. Furthermore, we can obtain the number of quanta in a frequency range $d\omega^{k}$

\begin{equation}
g(\omega)d\omega^{k}=\frac{V(\omega^{k})^{2}}{\pi^{2}c^{3}}d\omega^{k}\label{phase}.
\end{equation}

As both expressions Eq.\eqref{rj} and Eq.\eqref{phase} are related,  we conclude that under GUP modification they remain valid. In \cite{bose1994planck}, it is argued that the total number of cells must be considered equal to the number of
possible ways of placing a photon in the relevant volume. Then, we find the number of cells allowed for a range frequency $d\omega^{k}$

\begin{equation}
A^{k}=\frac{V(\omega^{k})^{2}}{\pi^{2}c^{3}}d\omega^{k}.   
\end{equation}

Making a connection with statistical thermodynamics, we define the entropy using the probability function $W$ from Eq.\eqref{eq11} that contains the information regarding the distribution of the cells

\begin{align}
S=k_{_{B}}\left\{\gamma E-\sum_{k} 3A^{k}\log(1-e^{-\gamma\hbar\omega^{k}})-A^{k}\log\left[(1-e^{-\gamma\hbar\omega^{k}})^{2}-2\gamma\beta^{k}e^{-\gamma\hbar\omega^{k}}\right]\right\}\label{s},
\end{align}

where $k_{_{B}}$ by the Boltzmann constant. By using the following thermodynamics relation $\frac{\partial S}{\partial E}=1/T$, we get the Lagrange multiplier $\gamma=1/k_{_{B}}T$.

From equation Eq.\eqref{eq8}, we can obtain the total energy 
\begin{align}
E=\sum_{k}\left[\hbar\omega^{k}+\beta^{k}(N^{k}+M^{k})\right]=V\int_{0}^{\infty}\rho_{_{T}}(\omega)\ d\omega,
\end{align}

where $\rho_{_{T}}(\omega)$ is the energy density. By performing the following transformation $\rho_{_{T}}(\lambda)=-\rho_{_{T}}(\omega)\frac{d\omega}{d\lambda}$, the energy density as a function of the wavelength is 

\begin{align}
& \rho_{_{T}}(\lambda)=\frac{8\pi{hc}}{\lambda^{5}}\Bigg\{\frac{1+\frac{hc}{\lambda}\gamma_{_{EM}}^{2}}{e^{\frac{hc}{\lambda k_{_{B}}T}}-1}\left[\frac{(e^{\frac{hc}{\lambda k_{_{B}}T}}-1)^{2}-2\frac{(hc\gamma_{_{EM}})^{2}}{\lambda^{2}k_{_{B}}T}e^{\frac{hc}{\lambda k_{_{B}}T}}(2+e^{\frac{hc}{\lambda k_{_{B}}T}})}{(e^{\frac{hc}{\lambda k_{_{B}}T}}-1)^{2}-2\frac{(hc\gamma_{_{EM}})^{2}}{\lambda^{2}k_{_{B}}T}e^{\frac{hc}{\lambda k_{_{B}}T}}}\right]\nonumber\\
&+ \frac{\frac{hc}{\lambda}\gamma_{_{EM}}^{2}}{(e^{\frac{hc}{\lambda k_{_{B}}T}}-1)^{2}}\left[\frac{(e^{\frac{hc}{\lambda k_{_{B}}T}}-1)^{2}(e^{\frac{hc}{\lambda k_{_{B}}T}}+1)-2\frac{(hc\gamma_{_{EM}})^{2}}{\lambda^{2}k_{_{B}}T}e^{\frac{hc}{\lambda k_{_{B}}T}}(4+7e^{\frac{hc}{\lambda k_{_{B}}T}}+e^{\frac{2hc}{\lambda k_{_{B}}T}})}{(e^{\frac{hc}{\lambda k_{_{B}}T}}-1)^{2}-2\frac{(hc\gamma_{_{EM}})^{2}}{\lambda^{2}k_{_{B}}T}e^{\frac{hc}{\lambda k_{_{B}}T}}}\right]\Bigg\}\label{u}.
\end{align}

It is worth noting that in the limit $\gamma_{_{EM}}\rightarrow 0$, the extra terms that carry the modification on the energy density vanish. In such a limit, $\rho_{_{T}}(\lambda)$ is reduced consistently to the usual expression 

\begin{align}
    \rho_{_{0T}}(\lambda)=\frac{8\pi\hbar c}{\lambda^{5}}\frac{1}{e^{\frac{hc}{\lambda k_{_{B}}T}}-1}.
\end{align}

\begin{figure}[]
\centering
\includegraphics[scale=1]{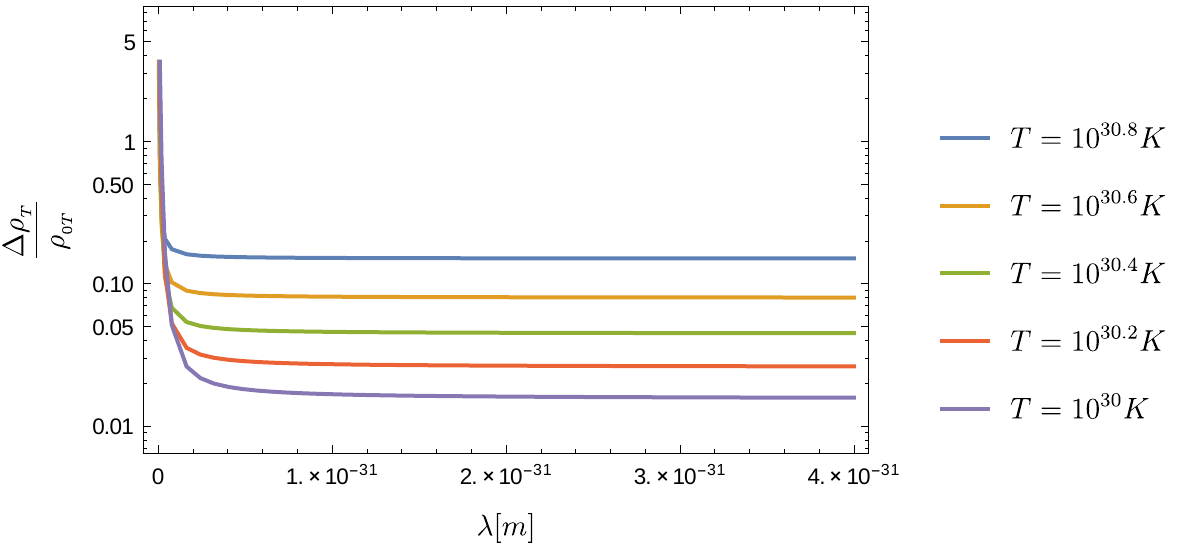}
\caption{Relative modification of the energy density $\frac{\Delta\rho_{_{T}}}{\rho_{_{0T}}}$ for different temperatures. For higher temperatures, the difference between the standard energy density and the modified one with GUP is larger. }
\end{figure}

We can study the modification in the energy density by plotting the relative modification 

\begin{equation}
\frac{\Delta\rho_{_{T}}}{\rho_{_{0T}}}=\frac{\rho_{_{0T}}-\rho_{_{T}}}{\rho_{_{0T}}}.    
\end{equation}

In Figure 1, we show the ratio for different temperatures. As it can be observed, the larger difference takes place at larger temperatures.

\section{Wien's Law}

In the black body distribution, Wien's law establishes a relation between a given temperature and the wavelength of the maximum of the distribution. In particular, the higher the temperature of the black body, the lower the wavelength of the maximum. As we are including a minimal length we expect a modification of this law at high temperatures. Wien's Law can be deduced by finding the maximum of the distribution in Eq.\eqref{u}. In the standard theory, that is, in the limit $\gamma_{_{EM}}\rightarrow 0$, this procedure leads to constant quantity $x=\frac{hc}{\lambda k_{_{B}}T}=5+W(0,-5e^{-5})$, where $W(z)$ is the Lambert $W$ function. In the present case, we consider the approximated expression

\begin{align}
    x=\frac{hc}{\lambda_{0}k_{_{B}}T}\left(1-\frac{\delta\lambda}{\lambda_{0}}\right)\label{eq23},
\end{align}

where $\lambda_{0}$ is the wavelength that satisfies Wien's law in the standard cases and $\delta\lambda$ is the shift on the wavelength of the maximum due to GUP.

In order to simplify the expression, we consider an expansion up to the first order in $k_{_{B}}T\gamma_{_{EM}}^{2}$. Such approximation is justified for temperatures much smaller than the Planck temperature. By differentiating Eq.\eqref{u} with respect to $\lambda$ and imposing the maximum condition, we get 

\begin{multline}
  2k_{_{B}}T\gamma_{_{EM}}^{2}xe^{x}\left[x^{2}(e^{2x}+4e^{x}+1)-8x(e^{2x}-1)+6(e^{x}-1)^{2}\right]\\
  -xe^{x}(e^{x}-1)^{2}+5(e^{x}-1)^{3}=0\label{wien}.
\end{multline}

Eq.\eqref{wien} can be solved numerically for $\frac{\delta\lambda}{\lambda_{0}}$ with the condition $\frac{\delta\lambda }{\lambda_{0}}<1$. In \(\text{Figure}\ 2\), we show the temperature dependence of relative shift for the wavelength of the maximum for different values of $\gamma_{_{0}}$. We observe that the modification grows with the temperature reaching the value $\frac{\delta\lambda}{\lambda_{0}}=1$. For such a value and beyond, the approximation in Eq.\eqref{eq23} cannot be considered valid. Specifically, for $\gamma_{_{0}}=1$, the approximation breaks close to the Planck temperature. Consistently with the approximation, the modification $\delta\lambda$ goes to $0$ for much smaller temperatures.

\begin{figure}[]
\centering
\includegraphics[scale=0.8]{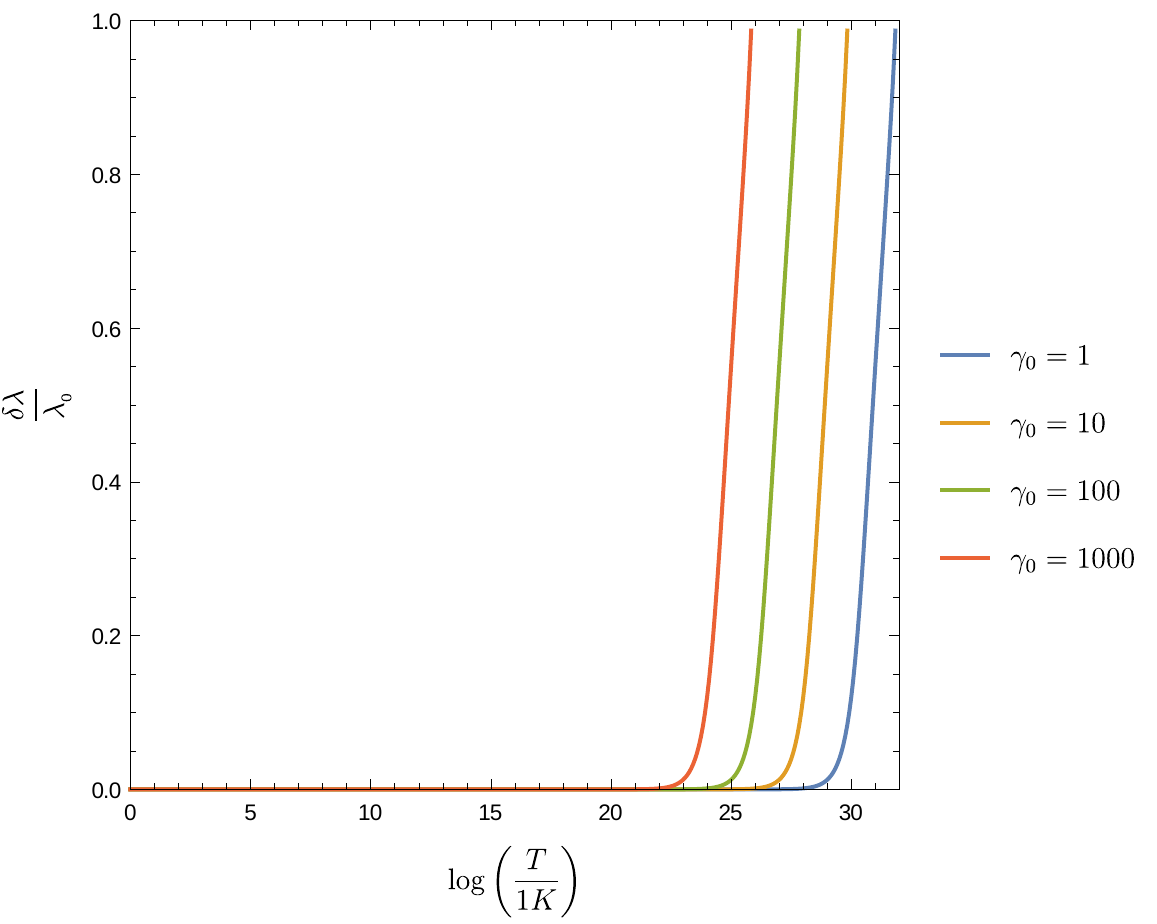}
\caption{Relative shift  $\frac{\delta\lambda}{\lambda_{0}}$ as a function of the temperature for different values of the parameter $\gamma_{_{0}}$. }
\end{figure}

\section{Stefan-Boltzmann Law}

The Stefan-Boltzmann law describes the total power radiated by a cavity with volume $V$ at an absolute temperature $T$. More precisely, the law establishes that the radiance of a black body, that is the amount of energy radiated per unit of surface is proportional to the fourth power of the absolute temperature

\begin{equation}
    R_{_{T}}=\sigma T^{4}\label{sb},
\end{equation}

where the proportionality constant, called the Stefan-Boltzmann constant, is 

\(\sigma=\frac{2\pi^{5}k_{_{B}}^{4}}{15h^{3}c^{2}}\).

This law can be derived from the total energy emitted by the black body by integrating  Eq.\eqref{u} and using the relation between  spectral radiance and the energy density $R_{_{T}}(\lambda)d\lambda=\frac{c}{4}\rho_{_{T}}(\lambda)d\lambda$. For simplicity, we expand Eq.\eqref{u} up to the first order in $k_{_{B}}T\gamma_{_{EM}}^{2}$.
Such approximation is justified for small values of the temperature compared to the Planck temperature. The total energy per unit volume is then  

\begin{equation}
\begin{split}
  R_{_{T}}&=\frac{2\pi^{5}k_{_{B}}^{4}}{15h^{3}c^{2}}T^{4}-\frac{4\pi k_{_{B}}^{4}}{h^{3}c^{2}}(k_{_{B}}T\gamma_{_{EM}}^{2})T^{4}\int_{0}^{\infty}\frac{x^{4}e^{x}(1+x-e^{x}+xe^{x})}{(e^{x}-1)^{3}}dx\\
 &=\frac{2\pi^{5}k_{_{B}}^{4}}{15h^{3}c^{2}}T^{4}(1-32k_{_{B}}T\gamma_{_{EM}}^{2}).
 \end{split}
\end{equation}

 By substituting the value of the Stefan-Boltzmann constant, we obtain the following equation

\begin{align}
    R_{_{T}}=\sigma T^{4}(1-32k_{_{B}}T\gamma_{_{EM}}^{2})\label{eq26}.
\end{align}

This is the first-order modification of the Stefan-Boltzmann law by including a minimal measurable length. We notice that in the limit $\gamma_{_{EM}}\rightarrow 0$, we recover the usual Stefan-Boltzmann law for the black body Eq.\eqref{sb}. 

In Figure 3, we plot the radiance for both the GUP modification and the ordinary case.

\begin{figure}[]
\centering
\includegraphics[scale=0.8]{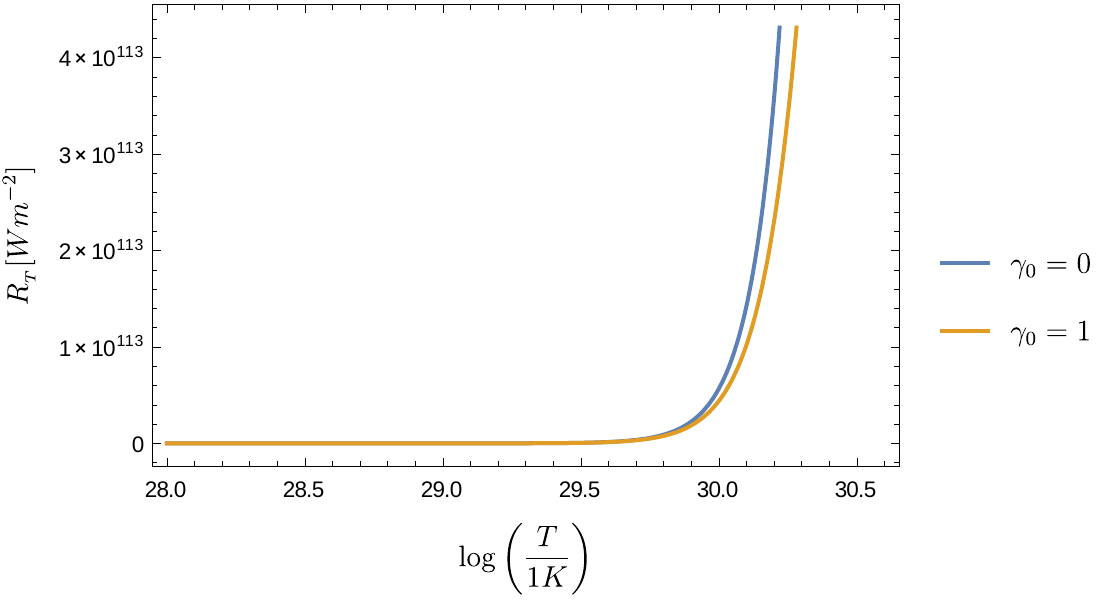}
\caption{Plot of the radiance of a black body $R_{_{T}}$. The solid blue line represents the Stefan-Boltzmann law in the ordinary theory. The solid orange line corresponds to the modified law in Eq.\eqref{eq26}}
\end{figure}

\section{Conclusions}
Statistical mechanics, as well as thermodynamics, may offer  indirect evidence of quantum gravity effects related to a minimal measurable length. The effects of such a length modifies the potentials as well as the laws established in both disciplines. However, the energy at which such effects become relevant is still outside the range offered by current experiments.\\ 
In this paper, we analyzed the effects of a minimal measurable length on the black body spectrum. To do so, we have considered a quantization procedure for the  electromagnetic field inspired by the GUP. One of the effects of such a procedure is that of modifying the dependence of the quanta of energy on the frequency. Using Bose's approach \cite{bose1994planck}, we obtained the Planck distribution that matches with the standard expression in the limit $\gamma_{_{EM}}\rightarrow 0$. The modified energy density at any given temperature results to be smaller than the energy density in the standard theory for the same temperature excluding values close to Planck length. Elaborating on Wien's law using the modified energy density, we found that GUP effects shift the maximum of the distribution.  We observed that such a modification depends on the temperature. For much smaller temperatures compared to the Planck temperature, the modification goes to zero, consistently with the approximation.   The modified Stefan-Boltzmann law was obtained by integrating the spectral radiance related to the modified energy density. The results suggest that the total energy radiated is lower than in the ordinary case at high temperatures. Such an effect is compatible with results obtained in DSR for a photon gas \cite{faruk2016planck}. For both modifications, the effects of a minimal measurable length are   temperature-dependent.\\
The importance of the black body radiation lies in its applications, in thermodynamics as well as other contexts. For example, the adsorption and emission of black holes make them similar to a black body \cite{lee1986black}. However, black holes cannot absorb wavelengths longer than their size \cite{anderson2002scattering}. Furthermore, \textit{gedanken} experiments in black hole thermodynamics consider black holes with the size of the order of Planck length. For such systems, the Hawking temperature is of the order of Planck temperature. At such a temperature, as we have seen in the paper, the Planck distribution is expected to be modified by quantum gravitational effects. Thus, the modification considered here may have a role in the thermodynamics of Planckian black holes. Furthermore, the study of black body radiation applies to cosmology as well \cite{dicke1965cosmic}. The  microwave background radiation is observed to be an almost perfect black body with a temperature of $2.7$ K \cite{white2002resource}. Thus, a modification in the Planck distribution due to the GUP can play a role in obtaining new information in the early stage of the universe by considering quantum gravity corrections.\\


\end{document}